\documentclass[11pt]{article}
\usepackage{amsfonts,epsfig}
\begin{document}
\vspace*{0cm}

\begin{center}
{\setlength{\baselineskip}{1.0cm}{ {\Large{\bf Rational extension and Jacobi-type {\boldmath{$X_m$}} solutions of a quantum nonlinear oscillator \\}} }}
\vspace*{1.0cm}
{\large{\sc{Axel Schulze-Halberg}}$^\ddagger$ and {
\\[1ex]
\sc{Barnana Roy}}$^\dagger$}
\end{center}
\noindent \\
$\ddagger$ Department of Mathematics and Actuarial Science and Department of Physics, Indiana University Northwest, 3400 Broadway,
Gary IN 46408, USA, e-mail: axgeschu@iun.edu, xbataxel@gmail.com
 \\ \\
$\dagger$ Physics \& Applied Mathematics Unit, Indian Statistical Institute, Kolkata 700108, India,
e-mail: barnana@isical.ac.in

\vspace*{1cm}
\begin{abstract} \vspace{.3cm}
\noindent
We construct a rational extension of a recently studied nonlinear quantum oscillator model. Our extended
model is shown to retain exact solvability, admitting a discrete spectrum and corresponding closed-form
solutions that are expressed through Jacobi-type $X_m$ exceptional orthogonal polynomials.

\end{abstract}

\section{Introduction}
This note is concerned with a one-dimensional model of a nonlinear quantum oscillator that has been
subject of several recent studies. This quantum model, along with its classical counterpart, was first
introduced and discussed in \cite{Mathews} \cite{mlquantum}. It presents two interesting features,
namely nonlinearity of the potential and the presence of a term that can be interpreted as a
position-dependent mass, producing a position-dependent spring constant in the classical oscillator potential.
This feature allows the existence of solutions for the classical nonlinear
oscillator that have a simple harmonic form. The quantum-mechanical version of
the model admits a discrete spectrum and a corresponding orthogonal
set of solutions with respect to a weighted Hilbert space. Both the spectral values and the solutions can
be obtained in closed form \cite{mlquantum} \cite{cari}. In particular, the solutions allow for a
representation in terms of special functions \cite{midyaosc} \cite{xbatjohn}.
The purpose of the present work is to present a particular generalization of the model and to show that
exact solvability is retained. More precisely, we will construct a rational extension of the nonlinear
oscillator potential, such that the associated solutions are expressed in terms of exceptional
orthogonal polynomials {\cite{Gomez1, Gomez2}.
 A complete set of such polynomials, introduced in \cite{ullate}, has the remarkable
property of forming an orthogonal system in a weighted Hilbert space, though they start with some polynomial of degree greater than
or equal to one.

A considerable amount of research has been dedicated to studying
exceptional orthogonal polynomials that lead to rational extensions
of quantum-mechanical potentials. There exist several approaches to obtain the latter, e.g. point canonical transformation
transformation \cite{quesne1, quesne, quesnemorse, midya1}, supersymmetric quantum mechanics \cite{quesne2, bagchi1, dutta,
 odake1, odake2}, Darboux-Crum transformation \cite{Gomez3, ullate},
 Darboux-Backlund transformation \cite{Grandati1, Grandati2, Grandati3} and prepotential approach \cite{Ho1,Ho2}.

 While most of the models are governed by Schr\"odinger-type equations, the nonlinear oscillator
system we focus on in this note obeys a slightly more general equation. By making use of a relation to the
well-known trigonometric Scarf problem \cite{dere} and its extended counterpart \cite{midya}, we are able to obtain a
rational extension of our nonlinear oscillator model, together with its discrete spectrum and the corresponding
solutions in explicit form. It should be mentioned that $X_m$ orthogonal polynomials are obtainable through several approaches In section 2 we will briefly summarize basic facts about the conventional nonlinear oscillator, while
section 3 is devoted to the construction of our rationally extended system and its discussion.

\section{Preliminaries}
We will now give a brief review of the nonlinear oscillator model, an extension of which will be constructed in the
remainder of this note. Besides introducing the model, we will comment on its interpretation as a
position-dependent mass system.

\paragraph{The nonlinear oscillator model.}
Let us first mention that a more detailed introduction of the system and discussion of further
properties can be found in \cite{mlquantum} and \cite{cari} \cite{xbatjohn}, respectively.
For $\lambda \in \mathbb{R}$, $\lambda<0$, define the interval
$D_\lambda \subset \mathbb{R}$ by
\begin{eqnarray}
D_\lambda &=& \left(0, |\lambda|^{-\frac{1}{2}} \hspace{.1cm}\right). \label{dlam}
\end{eqnarray}
We consider the following family of Dirichlet-type boundary-value problems
\begin{eqnarray}
(1-|\lambda|~x^2)~\Psi''-|\lambda|~x~\Psi'+\left[2~E-\frac{(1-|\lambda|)~x^2}{1-|\lambda|~x^2}\right] \Psi
&=&0,~~~x \in D_\lambda  \label{bvp1} \\[1ex]
\Psi\left(0 \right) ~~=~~ \Psi\left(|\lambda|^{-\frac{1}{2}} \right) &=& 0, \label{bvp2}
\end{eqnarray}
where $E \in \mathbb{R}$ represents the spectral parameter and the boundary conditions are allowed to be
understood in the sense of a limit. We refer to the potential $V$ of the above model as the function
\begin{eqnarray}
V &=& \frac{(1-|\lambda|)~x^2}{1-|\lambda|~x^2}. \label{conv}
\end{eqnarray}
Furthermore, we require $\Psi \in L_\mu^2\left(D_\lambda \right)$ for the
weight function $\mu$, given by
\begin{eqnarray}
\mu &=& \frac{1}{\sqrt{1-|\lambda|~x^2}}. \label{mu}
\end{eqnarray}
The problem (\ref{bvp1}), (\ref{bvp2}) admits a discrete
spectrum $(E_n)$ and
an infinite orthogonal set of corresponding solutions $(\Psi_n) \subset L_\mu^2\left(D_\lambda \right)$,
$n \in \mathbb{N} \cup \{0\}$, given by the following
expressions:
\begin{eqnarray}
E_n &=& \frac{|\lambda|}{2}~(2~n+1)^2+2~n+\frac{3}{2} \label{eneg} \\[1ex]
\Psi_n &=& \left(1-|\lambda|~x^2 \right)^\frac{1}{4} P_{-2n+\frac{1}{\lambda}-\frac{3}{2}}^{\frac{1}{\lambda}+
\frac{1}{2}} \left(-\sqrt{|\lambda|}~x \right), \label{psineg}
\end{eqnarray}
where $P$ stands for the associated Legendre function of the first kind \cite{abram}.

\paragraph{Interpretation as a positon-dependent mass system.}
The governing equation (\ref{bvp1}) of the nonlinear oscillator model can be obtained from the following
Hamiltonian:
\begin{eqnarray}
H &=&  \frac{1}{2}~\sqrt{\lambda~x^{2}+1}~p~\sqrt{\lambda~x^{2}+1}%
~p+V, \label{hampm}
\end{eqnarray}
where $V$ is given in (\ref{conv}) and $p=-i~d/dx$ stands for the conventional momentum operator. The
classical counterpart $H_c$ of (\ref{hampm}) reads
\begin{eqnarray}
H_c &=& \frac{1}{2}~(\lambda~x^{2}+1)~p^{2}+V, \label{champm}
\end{eqnarray}
note that here $p$ represents the classical momentum variable. The coefficient of $p$ can be interpreted
as a position-dependent mass $M$, given by
\begin{eqnarray}
M &=& \frac{1}{\lambda~x^{2}+1}. \label{pm}
\end{eqnarray}
The quantization of our system, that is, the transition from the classical function (\ref{champm}) to
the quantum operator (\ref{hampm}), is not unique. More precisely, there are several nonequivalent ways to
order the momentum operators and the term related to our position-dependent mass (\ref{pm}). This
ordering ambiguity is well-known in problems involving position-dependent mass systems \cite{roos},
where the general expression for a quantum Hamiltonian is given by
\begin{eqnarray}
H_{gen} &=& \frac{1}{4} \left(M^j p~M^k p~M^l+M^l p~M^k p~M^j\right)+V, \label{roos}
\end{eqnarray}
for a mass function $M$ and integers $j,k,l$ that must satisfy the relation $j+k+l=-1$. It is
interesting to observe that our
Hamiltonian (\ref{hampm}) for the mass function (\ref{pm}) cannot be written as a special
case of (\ref{roos}), no matter what parameter setting is employed (this can be verified by expanding both
(\ref{hampm}) and (\ref{roos})). This fact is most easily explained by observing that the Hamiltonian
$H_{gen}$ is defined on a conventional $L^2$ space (without weight), while our Hamiltonian is defined on a
weighted space $L_\mu^2$. Loosely speaking, the weight function (\ref{mu}) as part of the integral measure in
the inner product, cancels
out with the first root in (\ref{hampm}), such that the remaining terms in fact take the form (\ref{roos}) for
$j=l=0$, $k=-1$. In principle, we could choose an ordering of terms that is different from (\ref{hampm}) and that,
in general, would lead to a different governing equation. As an example we pick the well-known setting
$j=l=-1/2$ and $k=0$ in (\ref{roos}, leading to the equation \cite{zhu}
\begin{eqnarray}
(1-|\lambda|~x^2)~\Psi''-2~|\lambda|~x~\Psi'+\left[2~E-\frac{|\lambda|+2~(1-|\lambda|)~x^2}{1-|\lambda|~x^2}\right]
\Psi &=&0. \nonumber
\end{eqnarray}
We observe that this equation is different from its counterpart shown in (\ref{bvp1}). Another example for
a possible setting in (\ref{roos}) is given by $j=l=-1/4$ and $k=-1/2$. We obtain the following equation
\cite{mustafa}
\begin{eqnarray}
(1-|\lambda|~x^2)~\Psi''-2~|\lambda|~x~\Psi'+\left\{2~E-\frac{2~|\lambda|+[8-|\lambda|(|\lambda|+8)]~x^2}
{4-4~|\lambda|~x^2}\right\}
\Psi &=&0, \nonumber
\end{eqnarray}
which again is different from (\ref{bvp1}). It is clear that we can generate many more nonequivalent models
in a similar way, but we omit to consider them in detail here, as this would be beyond the scope of this note.

\section{The rationally extended nonlinear oscillator model}
We shall now construct a rationally extended version of the oscillator model (\ref{bvp1}), (\ref{bvp2}).
The key idea for this construction is the use of a relation between the latter model and the well-known
trigonometric Scarf system \cite{dere}. Since this relation has not been documented in the literature yet,
we will now introduce it and afterwards establish a link to the rational extensions.

\subsection{Nonlinear oscillator and trigonometric Scarf potential}
For $a,b \in \mathbb{R}$, $a,b>-1$, the following boundary-value problem is known as the trigonometric Scarf model:
\begin{eqnarray}
\Phi''+\left(\epsilon-\frac{2~a^2+2~b^2-1}{4}~\sec^2(u)+\frac{b^2-a^2}{2}~\sec(u)~\tan(u)\right) \Phi &=& 0,~~~|u|<\frac{\pi}{2}
\label{bvp3} \\
\Phi\left(-\frac{\pi}{2}\right) ~~=~~\Phi\left(\frac{\pi}{2}\right) &=& 0, \label{bvp4}
\end{eqnarray}
where $\epsilon \in \mathbb{R}$ stands for the spectral parameter. This problem admits a discrete spectrum and a corresponding orthogonal set of solutions in $L^2(-\pi/2,\pi/2)$,
the explicit form of which we do not need here \cite{dere}. Instead, we will now demonstrate that the above
boundary-value problem can be converted into our nonlinear oscillator model (\ref{bvp1}), (\ref{bvp2}). To this end,
we apply the following point transformation to the Schr\"odinger equation (\ref{bvp3}):
\begin{eqnarray}
u &=& -\frac{\pi}{2}+2~\mbox{arcsin}\left(\sqrt{|\lambda|}~x \right) \qquad \qquad \qquad \Psi(x) ~~=~~ \Phi(u(x)), \label{coord}
\end{eqnarray}
rendering the latter equation in the following form
\begin{eqnarray}
-\frac{1-|\lambda|~x^2}{4~|\lambda|}~\Psi''+\frac{1}{4}~x~\Psi'+\left[-\epsilon+\frac{4~b^2-1}{16~|\lambda|~x^2}+
\frac{4~a^2-1}{16~(1-|\lambda|~x^2)}\right] \Psi &=&0. \label{scarf1}
\end{eqnarray}
In the next step we redefine the parameters $a$, $b$ and $\epsilon$ as follows
\begin{eqnarray}
a ~~=~~ \frac{1}{2} \left| \frac{\lambda+2}{\lambda}
\right| \qquad \qquad \qquad b~~=~~\frac{1}{2} \qquad \qquad \qquad
\epsilon~~=~~\frac{1}{2~|\lambda|}~E+\frac{1}{4~\lambda^2}-\frac{1}{4~|\lambda|}. \label{pars}
\end{eqnarray}
Observe that these definitions are compatible with the domain of the parameters $a$ and $b$, which
are required to be larger than $-1$. Substitution of these settings converts equation (\ref{scarf1}) to
\begin{eqnarray}
-\frac{1-|\lambda|~x^2}{4~|\lambda|}~\Psi''+\frac{1}{4}~x~\Psi'+\left[-\frac{E}{2~|\lambda|}
+\frac{(1-|\lambda|)~x^2}{4~|\lambda|~(1-|\lambda|~x^2)} \right] \Psi &=&0. \label{osceq}
\end{eqnarray}
After multiplying this equation by $-4|\lambda|$ we obtain precisely equation (\ref{bvp1}). Next, let us verify that the domain $(-\pi/2,\pi/2)$ of equation
(\ref{bvp3}) transforms to $D_\lambda$, as given in (\ref{dlam}). On substituting $u=-\pi/2$ into the coordinate
change (\ref{coord}), we obtain after some elementary simplification
\begin{eqnarray}
0 &=& 2~\mbox{arcsin}\left(\sqrt{|\lambda|}~x \right). \label{b1}
\end{eqnarray}
The only real solution of this equation is given by $x=0$, coinciding with the left boundary point of $D_\lambda$.
In order to verify the right boundary point, let us now plug $u=\pi/2$ into (\ref{coord}). We arrive at the equation
\begin{eqnarray}
\pi &=& 2~\mbox{arcsin}\left(\sqrt{|\lambda|}~x \right), \label{b2}
\end{eqnarray}
the unique real solution of which is given by $x=|\lambda|^{-\frac{1}{2}}$, as desired. As a direct consequence of the
solutions to (\ref{b1}) and (\ref{b2}), we find that the boundary conditions (\ref{bvp4}) are mapped into their
counterparts (\ref{bvp2}) by means of our point transformation (\ref{coord}). In summary, we have shown that
the latter point transformation provides an interrelation between the nonlinear oscillator model
(\ref{bvp1}), (\ref{bvp2}) and the trigonometric Scarf system (\ref{bvp3}), (\ref{bvp4}).

\subsection{Construction of the rational extension}
The method of constructing a rational extension of our nonlinear oscillator model (\ref{bvp1}), (\ref{bvp2}) is
straightforward. First we observe that there is a rational extension of the trigonometric Scarf system
(\ref{bvp3}), (\ref{bvp4}) that has been studied previously, see e.g. \cite{midya} \cite{quesne}. This means that by
simply applying our point transformation (\ref{coord}) to the extended Scarf system, we will obtain the sought
oscillator counterpart. To make this note self-contained, we quote the main results for rationally extended Scarf system.

\paragraph{The rationally extended Scarf system.}
Let $a,b \in \mathbb{R}$, $a,b>-1$. We consider the model governed by the following boundary-value problem
\begin{eqnarray}
\hspace{-.3cm}\Phi''+\left(\epsilon-\frac{2~a^2+2~b^2-1}{4}~\sec^2(u)+\frac{b^2-a^2}{2}~\sec(u)~\tan(u) + r(u)\right) \Phi &=& 0,
~ |u|<\frac{\pi}{2}
\label{bvpe1} \\[1ex]
\Phi\left(-\frac{\pi}{2}\right) ~~=~~\Phi\left(\frac{\pi}{2}\right) &=& 0. \label{bvpe2}
\end{eqnarray}
As before, $\epsilon$ denotes the
spectral parameter, and the function $r$ is given for an $m \in \mathbb{N}$ by
\begin{eqnarray}
r(u) &=& -2~m~(a-b-m+1)-(a-b-m+1)\left[a+b+(a-b+1)~\sin(u)\right] \times \nonumber \\[1ex]
&\times&
\frac{P_{m-1}^{(-a,b)}[\sin(u)]}{P_{m}^{(-a-1,b-1)}[\sin(u)]}+
\frac{(a-b-m+1)^2}{2}~\cos^2(u) \left\{
\frac{P_{m-1}^{(-a,b)}[\sin(u)]}{P_{m}^{(-a-1,b-1)}[\sin(u)]}
\right\}^2, \label{r}
\end{eqnarray}
where $P$ denotes a Jacobi polynomial \cite{abram}.
The problem (\ref{bvpe1})-(\ref{r}) is known as the rationally extended Scarf model \cite{midya} \cite{quesne}, if the
following additional constraints on the parameters $a,b$ and $m$ are imposed:
\begin{eqnarray}
\begin{array}{ll}
b ~\neq~ 0 \qquad \qquad & a,~a-b-m+1 ~\notin~ \{0,1,...,m-1\} \\
a-m+2 ~>~ 0 \qquad \qquad & \mbox{sign}(a-m+1)-
\mbox{sign}(b) ~=~ 0.
\end{array} \label{ullatep}
\end{eqnarray}
This boundary value problem
(\ref{bvpe1})-(\ref{r}) admits
an infinite discrete spectrum $(\epsilon_n)$ and corresponding solutions $(\Phi_n)$,
$n \in \mathbb{N} \cup \{0\}$, $n \geq m$. The
explicit form of these objects reads
\begin{eqnarray}
\epsilon_n &=& \left(\frac{2~n-2~m+a+b+1}{2}\right)^2 \label{enescarf} \\[1ex]
\Phi_n&=& \frac{\left[1-\sin(u)\right]^{\frac{a}{2}+\frac{1}{4}} \left[1+\sin(u)\right]^{\frac{b}{2}+\frac{1}{4}}}
{P_m^{(-a-1,b-1)}[\sin(u)]} ~{\cal P}_n^{(a,b,m)}[\sin(u)]. \label{solscarf}
\end{eqnarray}
Observe that the symbol ${\cal P}$ represents an exceptional polynomial of $X_m$ Jacobi type. The class of
such polynomials are related to classical Jacobi polynomials as \cite{midya}
\begin{eqnarray}
{\cal P}_n^{(a,b,m)}(x) &=& (-1)^m \left[\frac{a+b+n-m+1}{2~(a+n-m+1)}~(x-1)~P_m^{(-a-1,b-1)}(x)~
P_{n-m-1}^{(a+2,b)}(x)+ \right. \nonumber \\[1ex]
&+& \left.\frac{a-m+1}{a+n-m+1}~P_m^{(-a-2,b)}(x)~P_{n-m}^{(a+1,b-1)}(x)\right], \nonumber
\end{eqnarray}
recall that $n \geq m$. It is important to note that the conditions (19) guarantee that the denominator in (21) does not
have any zeroes inside or at the endpoints of the interval $(-\frac{\pi}{2}, \frac{\pi}{2})$, such that the solutions
(21) are free of singularities, and as such can be normalized.\\
The solution set (\ref{solscarf}) forms an orthogonal set with respect to the Hilbert
space $L^2(-\pi/2,\pi/2)$, that is, for $n,l \in \mathbb{N}$, $n,l \geq m$,
\begin{eqnarray}
\int\limits_{(-\frac{\pi}{2},\frac{\pi}{2})} \Phi^\ast_l~\Phi_n~du &=& N~\delta_{ln}, \label{inner}
\end{eqnarray}
where the asterisk denotes complex conjugation and
the normalization constant $N \in \mathbb{R}$ is explicitly given by \cite{midya}
\begin{eqnarray}
N &=& (-1)^m~\pi~2^{a+b+1}~(n+b)~(n-2~m+a+1)~\Gamma(n-m+a+2) \times \nonumber \\[1ex]
&\times& \bigg\{
\sin\left[(n+b)~\pi\right] ~(2~n-2~m+a+b+1)~(n-m+a+1)~\Gamma(n-m+a+b+1) \times \nonumber \\
&\times& \Gamma(m-n-b+1)~
\Gamma(n-m+1) \bigg\}^{-1}. \label{n}
\end{eqnarray}
Note that $\Gamma$ stands for the Gamma function \cite{abram}.

\paragraph{Transformation to the rationally extended oscillator system.}
In order to obtain a rational extension of our nonlinear oscillator model (\ref{bvp1}), (\ref{bvp2}), we apply the
point transformation (\ref{coord}) and the parameter redefinition (\ref{pars}) to the Schr\"odinger equation
(\ref{bvpe1}) of the extended Scarf system. The resulting equation is given by
\begin{eqnarray}
-\frac{1-|\lambda|~x^2}{4~|\lambda|}~\Psi''+\frac{1}{4}~x~\Psi'+\left[-\frac{E}{2~|\lambda|}
+\frac{(1-|\lambda|)~x^2}{4~|\lambda|~(1-|\lambda|~x^2)}+R \right] \Psi &=&0, \label{osceq1}
\end{eqnarray}
where the function $R$ is defined through its counterpart $r$ from (\ref{r}) as follows
\begin{eqnarray}
R &=& r\left[-\frac{\pi}{2}+2~\mbox{arcsin}\left(\sqrt{|\lambda|}~x \right) \right]. \label{rx}
\end{eqnarray}
Before we state the explicit form of $R$ and the extended oscillator potential, let us observe that
(\ref{osceq1}) is very similar to the non-extended case (\ref{osceq}), as expected. Multiplication of
(\ref{osceq1}) by $-4~|\lambda|$ renders the latter equation in its final form. Together with the transformed
boundary condition, the rationally extended oscillator model (\ref{bvp1}), (\ref{bvp2}) then reads
\begin{eqnarray}
\left(1-|\lambda|~x^2\right)\Psi''-|\lambda|~x~\Psi'+\left[2~E
-\frac{(1-|\lambda|)~x^2}{1-|\lambda|~x^2}-4~|\lambda|~R \right] \Psi &=&0,~~~x \in D_\lambda \label{bvpx1} \\[1ex]
\Psi\left(0 \right) ~~=~~ \Psi\left(|\lambda|^{-\frac{1}{2}} \right) &=& 0, \label{bvpx2}
\end{eqnarray}
We will now state the extended potential associated with the latter problem. Note that the explicit form of $R$ is obtained
from (\ref{r}), evaluated at the coordinate $u$, as given in (\ref{coord}). We get the extended potential in the form
\begin{eqnarray}
V_{{\footnotesize{\mbox{ext}}}} &=& \frac{(1-|\lambda|)~x^2}{1-|\lambda|~x^2}+4~|\lambda|~R \nonumber \\[1ex]
&=&
\frac{(1-|\lambda|)~x^2}{1-|\lambda|~x^2}+4~m~|\lambda| \left(1-2~m+\left|\frac{\lambda+2}{\lambda} \right| \right)
- \nonumber \\[1ex]
&-& ~2~\lambda^2~x^2 \left(1-|\lambda|~x^2 \right) \left(1-2~m+\left|\frac{\lambda+2}{\lambda} \right| \right)^2
\left[
\frac{P_{m-1}^{\left(-\frac{1}{2}~\left|\frac{\lambda+2}{\lambda} \right|,\frac{1}{2}\right)}\left(2~|\lambda|~x^2-1\right)}
{P_{m}^{\left(-1-\frac{1}{2}~\left|\frac{\lambda+2}{\lambda} \right|,-\frac{1}{2}\right)}\left(2~|\lambda|~x^2-1\right)}
\right]^2 +\nonumber \\
& +&~2~\lambda^2~x^2~\left(1+\left|\frac{\lambda+2}{\lambda} \right|\right)
\left(1-2~m+\left|\frac{\lambda+2}{\lambda} \right| \right)
\frac{P_{m-1}^{\left(-\frac{1}{2}~\left|\frac{\lambda+2}{\lambda} \right|,\frac{1}{2}\right)}\left(2~|\lambda|~x^2-1\right)}
{P_{m}^{\left(-1-\frac{1}{2}~\left|\frac{\lambda+2}{\lambda} \right|,-\frac{1}{2}\right)}\left(2~|\lambda|~x^2-1\right)}.
\nonumber \\ \label{bvpx3}
\end{eqnarray}
Note that the function $R$ vanishes for $m=0$, such that the conventional oscillator problem (\ref{bvp1}), (\ref{bvp2}) is
recovered. Before we study the potential $V_{{\footnotesize{\mbox{ext}}}}$ in more detail and discuss its regularity, let us first construct the discrete spectrum and the solutions to the
boundary-value problem (\ref{bvpx1}), (\ref{bvpx2}). Starting out with the spectrum, we observe that the
spectral values $E_n$, $n \in \mathbb{N} \cup \{0\}$, can be constructed by solving the rightmost
relation in (12) for $E$ and substituting (\ref{enescarf})
for $\epsilon$. This calculation yields the following result, recall that $n\geq m$,
\begin{eqnarray}
E_n &=& \frac{1}{2}~|\lambda|~(2~m-2~n-1)^2+\frac{1}{2}~(4~n-4~m+3). \label{eneext}
\end{eqnarray}
As in the case of the extended potential (\ref{bvpx3}), we can easily see that
for $m=0$ the standard case (\ref{eneg}) is recovered. We emphasize that the spectral values (\ref{eneext}),
as well as the solutions that we will state in the following, are only
valid under certain restrictions to be imposed on the parameters $\lambda,~m$ and $n$, similar to (\ref{ullatep})
for the Scarf system. For the sake of simplicity, we will discuss these parameter restrictions at the end of this
section. Now, the solutions corresponding to the spectral values (\ref{eneext}) are obtained by applying the
point transformation (\ref{coord}) to the solutions (\ref{solscarf}) of the extended Scarf system. We obtain
the following infinite set of functions $\Psi_n(x) = \Phi_n(u(x))$, $n \in \mathbb{N} \cup \{0\}$, of the form
\begin{eqnarray}
\Psi_n &=& \frac{x~(2-2~x^2~|\lambda|)^{\frac{1}{4}+\frac{1}{4} \left|\frac{\lambda+2}{\lambda} \right|}}
{P_{m}^{\left(-1-\frac{1}{2}\left|\frac{\lambda+2}{\lambda} \right|,-\frac{1}{2}\right)}\left(2~|\lambda|~x^2-1\right)}~~
{\cal P}_{n}^{\left(\frac{1}{2}\left|\frac{\lambda+2}{\lambda} \right|,~\frac{1}{2},~m\right)}\left(2~|\lambda|~x^2-1\right), \label{solx}
\end{eqnarray}
where some irrelevant factors were omitted. These functions, which turn into (\ref{psineg}) for $m=0$, form an orthogonal set in the weighted
Hilbert space $L^2_\mu\left(D_\lambda \right)$, the
weight $\mu$ of which will be determined now. Taking into account the coordinate change in (\ref{coord}), the relation
between the differentials $du$ and $dx$ is found to be
\begin{eqnarray}
du &=& \frac{2~|\lambda|}{\sqrt{1-|\lambda|~x^2}}~dx. \nonumber
\end{eqnarray}
If we discard the irrelevant numerator of the fraction on the right-hand side, our weight function $\mu$ coincides
with its conventional counterpart (\ref{mu}). Therefore, the solutions (\ref{solx}) of the extended oscillator
system satisfy the following orthogonality relation for $n,l \in \mathbb{N}$, $n,l \geq m$,
\begin{eqnarray}
\int\limits_{D_\lambda} \Psi_l^\ast~\Psi_n ~\frac{1}{\sqrt{1-|\lambda|~x^2}}~dx &=& N~\delta_{ln}. \label{iprod}
\end{eqnarray}
The normalization constant $N$ can be derived from (\ref{n}) by simply inserting the settings (\ref{pars}) for the
parameters $a$ and $b$:
\begin{eqnarray}
N &=& (-1)^m~2^{\frac{1}{|\lambda|}}~\pi~(2~n+1)~|\lambda|~[2+|\lambda|~(2~n-4~m+1)]~
\Gamma\left(n-m+\frac{1}{|\lambda|}+\frac{1}{2}\right) \times \nonumber \\[1ex]
&\times&\bigg\{ \cos(n~\pi) ~[1+|\lambda| ~(2~n-2~m+1)]~[2+|\lambda|~(2~n-2~m+1) ]~
\Gamma\left(m-n+\frac{1}{2}\right) \times \nonumber \\[1ex]
&\times& \Gamma\left(n-m+1 \right) \Gamma\left(n-m+\frac{1}{|\lambda|}+1\right)
\bigg\}^{-1}. \nonumber
\end{eqnarray}

\paragraph{Regularity and constraints on the parameters.}
We recall that our extended Scarf system obeys the parameter conditions (\ref{ullatep}). In the present
paragraph, we will determine a corresponding set of conditions, adjusted to the new parameter $\lambda$ in the nonlinear
oscillator model. In the first step, we take a look at the denominator in (\ref{solx}), which is a Jacobi polynomial.
According to \cite{szego}, the latter polynomial vanishes at $x=0$, if and only if its second argument in the exponent
is a negative integer netween $-m$ and $-1$. Since this is not fulfilled due to the exponent being constant,
the denominator in (\ref{solx}) does never turn zero at $x=0$. The
second endpoint $x=|\lambda|^{-1/2}$ of $D_\lambda$ is a zero of the Jacobi polynomial in the denominator of
(\ref{solx}), if and only if the first argument in the exponent is a negative integer between $-m$ and $-1$ \cite{szego}.
This condition is satisfied, if for an $L \in \mathbb{N}$, $L\leq m$, the parameter $\lambda$ obeys
\begin{eqnarray}
1+\frac{1}{2}\left|\frac{\lambda+2}{\lambda} \right| &=& L. \label{lameq}
\end{eqnarray}
Therefore, in order to guarantee that the denominator of (\ref{solx}) does not vanish, we must have
\begin{eqnarray}
|\lambda| &\neq& \frac{2}{2~L-1},~~~L \in \mathbb{N},~ L\leq m. \label{lamsing1}
\end{eqnarray}
Observe that equation (\ref{lameq}) has more solutions, but they all lead to positive $\lambda$, which is not
allowed due to our standing assumption $\lambda<0$. In the last step we consider the interior of
$D_\lambda$, applicable results for which in \cite{ullate} give the restrictions stated in the second line of (\ref{ullatep}).
After incorporation of the settings (\ref{pars}) for $a$ and $b$ and their evaluation, we obtain that the
denominator of (\ref{solx}) does not have any zeros in $D_\lambda$, if and only if the following condition is
fulfilled
\begin{eqnarray}
\begin{array}{lll}
\lambda ~<~0 & \mbox{if} & m=1 \\[1ex]
{\displaystyle{\lambda \in \left(\frac{2}{1-2~m}~,~0 \right)}} & \mbox{if} & m>1.
\end{array} \label{lamsing2}
\end{eqnarray}
We observe that the denominator in (\ref{solx}) is the only factor of the solutions that can contribute a singularity.
Therefore, if both conditions (\ref{lamsing1}) and (\ref{lamsing2}) are satisfied simultaneously, the solutions
(\ref{solx}) are regular on the whole domain $D_\lambda$. In principle, it is not clear that regularity of the solutions
implies their normalizability, because the weight function (\ref{mu}) contributes a power $-1/2$ at the right
endpoint $x=|\lambda|^{-1/2}$ of $D_\lambda$, as can be seen from the inner product (\ref{iprod}). However,
assuming that the conditions (\ref{lamsing1}), (\ref{lamsing2}) hold, the overall exponent of $|\Psi_n|^2 \mu$
at $x=|\lambda|^{-1/2}$ is given by
\begin{eqnarray}
\frac{1}{2}+\frac{1}{2}\left|\frac{\lambda+2}{\lambda} \right| -\frac{1}{2}&=&
\frac{1}{2}\left|\frac{\lambda+2}{\lambda} \right|. \label{ppos}
\end{eqnarray}
Note that the last term on the left-hand side is contributed by the weight function (\ref{mu}). Since the
last expression in (\ref{ppos}) cannot attain negative values, we have shown that (\ref{lamsing1}), (\ref{lamsing2})
guarantee normalizability in the weighted Hilbert space $L^2_\mu\left(D_\lambda \right)$. After having discussed
regularity of the solutions associated with the extended oscillator potential, it remains to discuss regularity of the
latter potential itself. Inspection of its explicit form (\ref{bvpx3}) shows the following:
\begin{itemize}
\item The first term in the potential, which is contributed from the conventional system (\ref{bvp1}), (\ref{bvp2}),
generates a singularity at the right endpoint $x=|\lambda|^{-1/2}$ of the domain $D_\lambda$. This
singularity is inherent to the model and as such does not represent an issue.
\item Besides the term stemming from (\ref{conv}), singularities can only be generated by
the two denominator terms consisting of Jacobi polynomials. We observe that these polynomials coincide
precisely with their counterpart in the solutions (\ref{solx}). Consequently, fulfilling our conditions
(\ref{lamsing1}) and (\ref{lamsing2}) guarantees that the extended oscillator potential (\ref{bvpx3})
does not have singularities in the interval $D_\lambda$ or at its endpoints, except for the one
produced by the first term.
\end{itemize}
As we have now completed our discussion on regularity of the extended potential and its associated solutions,
we will proceed with the presentation of some applications.

\subsection{Special cases and examples}
We will now present a few particular cases of our extended nonlinear oscillator potential, along with its
corresponding solutions and spectral values. One particular point of emphasis will be the comparison between
the conventional model (\ref{bvp1}), (\ref{bvp2}) and its extended counterpart (\ref{bvpx1})-(\ref{bvpx3}).
\paragraph{The extended nonlinear oscillator potential.}
Starting out with the extended potential, we first recall that the latter function is given explicitly in (\ref{bvpx3}).\\ \\
{\bf Case 1.}
The potential (28) together with its conventional partner are shown in figure 1 for the parameter settings $m=1$ and $\lambda=-1/20$.
 Evaluation of (\ref{bvpx3}) for the present settings yields
\begin{eqnarray}
V_{{\footnotesize{\mbox{ext}}}} &=& \frac{\frac{19}{20}~x^2}{1-\frac{1}{20} ~x^2}
+\frac{80}{10+19~x^2}-\frac{1560}{\left(10+19~x^2 \right)^2}. \nonumber
\end{eqnarray}
The standard potential (\ref{conv}) is precisely the first term on the right-hand side, as required.
Since $\lambda$ is small in absolute value, the function (\ref{conv}) is close
to the harmonic oscillator potential, as can be seen in figure 1.
\begin{figure}[h]
\begin{center}
\epsfig{file=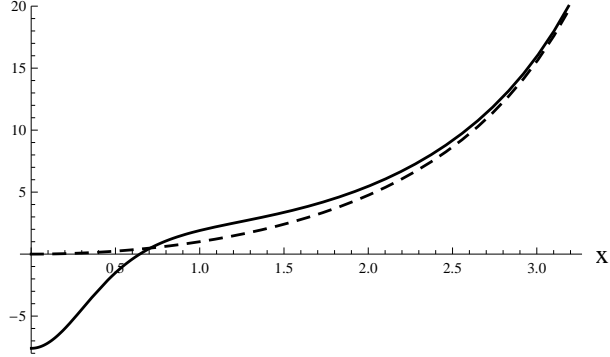,width=8cm}
\caption{The extended oscillator potential (\ref{bvpx3}) (solid line), compared to the conventional case (\ref{conv})
(dashed line) for the parameter settings $m=1$ and $\lambda=-1/20$.}
\end{center}
\end{figure}
\noindent \\
{\bf Case 2.} Here we take $m=1$, $\lambda = -5$. Note
that according to (\ref{lamsing2}) this cannot be done for any $m>1$, because $\lambda$ needs to always
be greater than $-2/3$. In this case the extended potential (28) is the following

\begin{eqnarray}
V_{{\footnotesize{\mbox{ext}}}} &=& -\frac{4~x^2}{1-5~x^2}+\frac{32}{1-2~x^2}-
\frac{24}{\left(2~x^2-1\right)^2}. \label{ex2}
\end{eqnarray}
This potential, together with (\ref{conv}), is displayed in figure 2. As before, we can identify the first term of
(\ref{ex2}) as the contribution from (\ref{conv}).
\begin{figure}[h]
\begin{center}
\epsfig{file=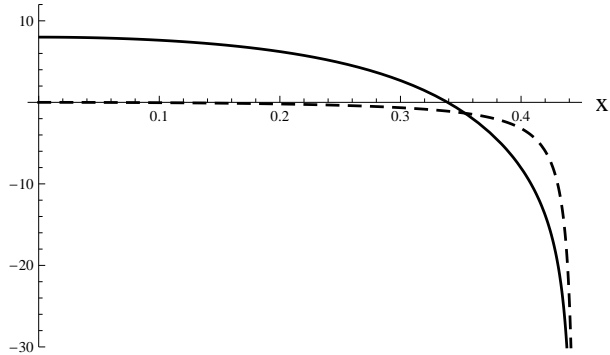,width=8cm}
\caption{The extended oscillator potential (\ref{bvpx3}) (solid line), compared to the conventional case (\ref{conv})
(dashed line) for the parameter settings $m=1$ and $\lambda=-5$.}
\end{center}
\end{figure}
We see that due to the negative sign in front of the leading term in (\ref{ex2}),
both potentials tend towards negative infinity close to the right endpoint $x=\sqrt{1/5}$ of the domain $D_{-5}$.\\ \\
{\bf Case 3.} The parameter settings in this case are $m=2$ and $\lambda=-13/20$. Observe that
the latter value of $\lambda$ is greater than $-2/3$ and as such remains consistent with our
regularity condition (\ref{lamsing2}). The extended potential can be extracted from (\ref{bvpx3}) as
\begin{eqnarray}
V_{{\footnotesize{\mbox{ext}}}} &=& \frac{\frac{7}{20}~x^2}{1-\frac{13}{20}~x^2} -
\frac{160~(-35+19~x^2)}{19~(50-60~x^2+19~x^4)}
+\frac{4000~(-13+8~x^2)}{19~\left(50-60x^2+19~x^4 \right)^2}. \nonumber
\end{eqnarray}
We see that the first term of the latter expression coincides with (\ref{conv}) for the present settings. The extended
potential and its conventional counterpart are shown in figure 3.
\begin{figure}[h]
\begin{center}
\epsfig{file=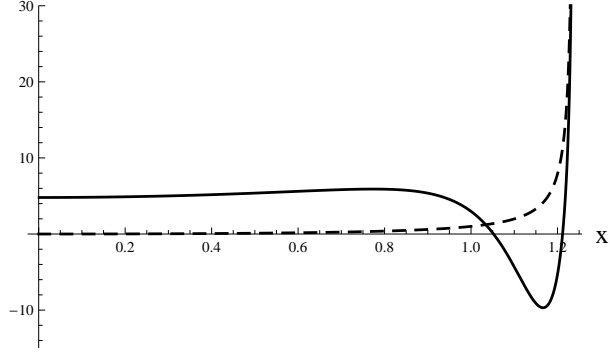,width=8cm}
\caption{The extended oscillator potential (\ref{bvpx3}) (solid line), compared to the conventional case (\ref{conv})
(dashed line) for the parameter settings $m=2$ and $\lambda=-13/20$.}
\end{center}
\end{figure}

\paragraph{Solutions to the extended oscillator model.}
In this paragraph we will study the solutions (\ref{solx}) for our extended nonlinear oscillator system
(\ref{bvpx1})-(\ref{bvpx3}) for several parameter settings. These solutions and their associated spectra
(\ref{eneext}) shall be compared to the conventional counterparts (\ref{psineg}) and (\ref{eneg}), respectively.
In our first example we consider solutions associated with the lowest spectral value (ground states). Figure 4
shows the probability density $|\Psi_0|^2 \mu$ of the conventional solution (\ref{psineg}) for $n=0$,
compared to the ground state probability densities $|\Psi_j|^2 \mu$, $j=1,2,3$, for the cases $m=1,2,3$ of the
extended model (\ref{solx}), respectively. We point out that the solutions associated with the
probability densities shown in figure 4 have been normalized. Note further
that only odd-numbered parameters $m$ are used in order to maintain legibility of the figure.
The solutions (\ref{solx}) of our extended model, probability densities of which are plotted in figure 4, have the
following explicit unnormalized form
\begin{eqnarray}
\Psi_0 &=& x~(x^2-10)^5 \nonumber \\
\Psi_1 &=& \frac{x~(x^2-10)^5~(3~x^2+5)}{(9~x^2+5)} \nonumber \\[1ex]
\Psi_3 &=& \frac{x~(x^2-10)^5~(2x^6+42~x^4+175~x^2+125)}{[(2~x^4+30~x^2+75)~7~x^2+125]} \nonumber \\[1ex]
\Psi_5 &=& \frac{x~(x^2-10)^5~\{[(x^6+275~x^4+12375~x^2+144375)8~x^2+3609375]~x^2+2165625\}}
{\{[(x^6+225~x^4+7875~x^2+65625)~8~x^2+984375]~x^2+196875\}}. \nonumber \\
\end{eqnarray}
\begin{figure}[h]
\begin{center}
\epsfig{file=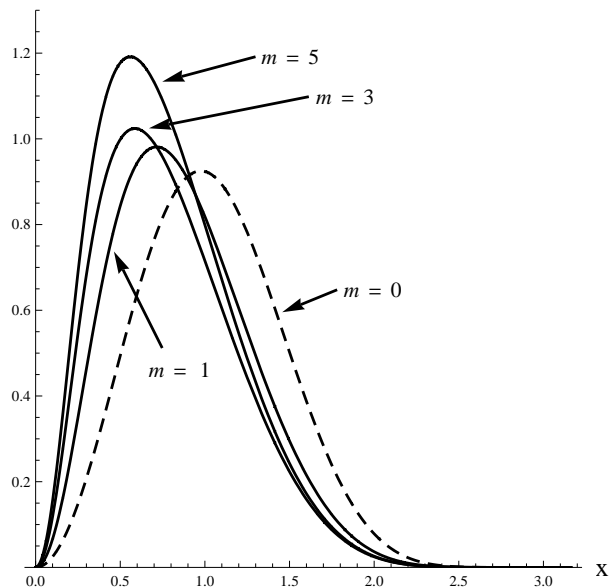,width=8cm}
\caption{Probability densities from the solutions (\ref{solx}) to the extended model (solid lines)
for $\lambda=-1/10$ and several values of $m=n$, compared to the conventional
probability density from (\ref{psineg}) (dashed line) for $m=n=0$. }
\end{center}
\end{figure}
All these solutions belong to the same spectral value, as evaluation of (\ref{eneext}) for $n=m$ shows:
\begin{eqnarray}
E_m &=& \frac{1}{20}+\frac{3}{2} ~~=~~ \frac{31}{20}. \nonumber
\end{eqnarray}
Recall that the conventional case (\ref{eneg}) is included here, as it arises from the extended model for $m=0$.
In the next example we examine probability densities of the extended model for the settings $m=3$ and
$\lambda=-1/5$, note that these values are compatible with our regularity condition (\ref{lamsing2}). Figure 5
shows a comparison of the latter probability densities with their conventional counterparts. As can be inferred
from the number of zeros that the probability densities exhibit, solutions belonging to the three lowest
spectral values have been used.
\begin{figure}[h]
\begin{center}
\epsfig{file=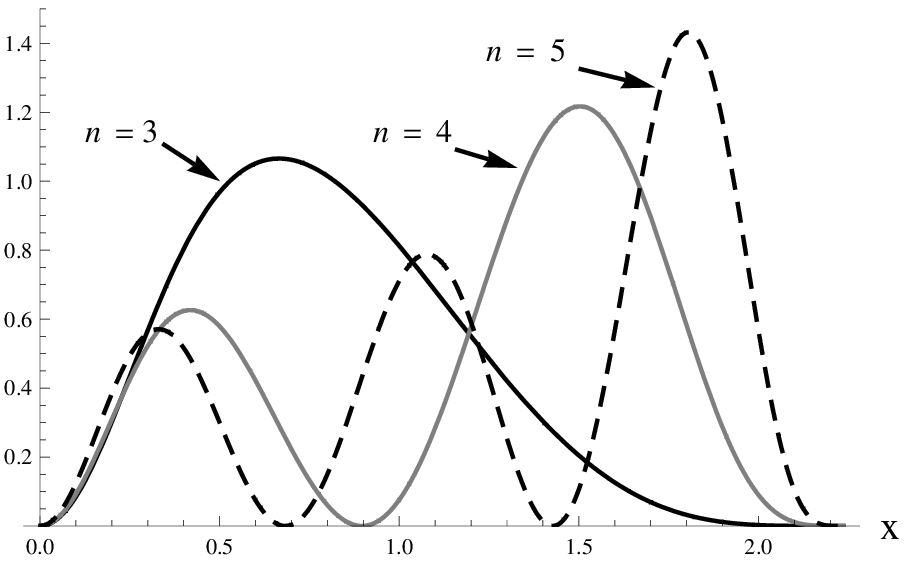,width=7cm} \hspace{.4cm}
\epsfig{file=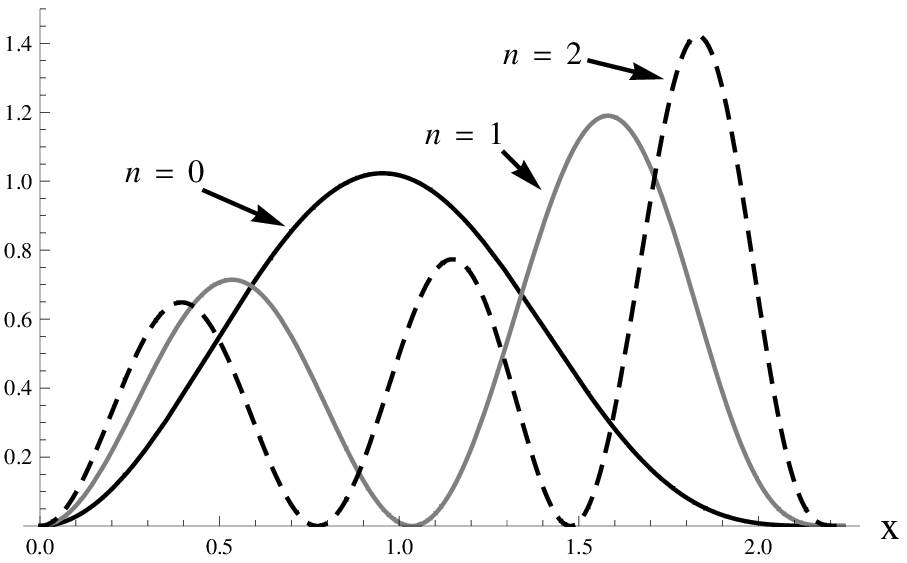,width=7cm}
\caption{Probability densities from the solutions (\ref{solx}) to the extended model
for $\lambda=-1/5$ and values $n=3,4,5$ (left plot), compared to the same situation for the conventional
probability density from (\ref{psineg}) (right plot), where we have $n=0,1,2$.}
\end{center}
\end{figure}
These spectral values for the extended case (\ref{eneext}) are given numerically by
\begin{eqnarray}
E_3~=~\frac{8}{5} \qquad E_4 ~=~ \frac{22}{5} \qquad E_5~=~8. \nonumber
\end{eqnarray}
They coincide with their respective partners (\ref{eneg}) in the conventional model for $n=0,1,2$.
Recall that in our extended model the lowest spectral value (\ref{eneext}) is
obtained for $n=m$.

\section{Concluding remarks}
In the present note we have constructed an exactly-solvable rational extension of the nonlinear oscillator
model (\ref{bvp1}), (\ref{bvp2}). Our construction makes use of a point transformation that interrelates the latter
model to the Scarf system, a rational extension of which is well-known. It is interesting to note that our
initial model, as well as its extended counterpart, are governed by an equation that is not of conventional
Schr\"odinger type, but involves a first-derivative term. Therefore, modifications of our method can be used
to obtain rational extensions of models more general than the Schr\"odinger context.

\end{document}